# Defect Engineered Room-Temperature Ferromagnetism in Quasi-Two-Dimensional Nitrided CoTa$_2$O$_6$

Yalin Ma,[1] Shuang Zhao,[1] Xiao Zhou,[1] Yijie Zeng,[2,3,*] Haili Song,[1] Jing Wang,[1] Guangqin Li,[1] Corey E. Frank,[4] Lu Ma,[5] Mark Croft,[6] Yonggang Wang,[7] Martha Greenblatt,[4] Dao-Xin Yao,[2,†] and Man-Rong Li[1,‡]

[1] *Key Laboratory of Bioinorganic and Synthetic Chemistry of Ministry of Education, School of Chemistry, Sun Yat-Sen University, Guangzhou 510275, China.*

[2] *State Key Laboratory of Optoelectronic Materials and Technologies, School of Physics, Sun Yat-Sen University, Guangzhou 510275, China.*

[3] *College of science, Hangzhou Dianzi University, Hangzhou 310018, China.*

[4] *Department of Chemistry and Chemical Biology, Rutgers, The State University of New Jersey, 123 Bevier Road, Piscataway, New Jersey 08854, USA*

[5] *Brookhaven National Laboratory, National Synchrotron Light Source II, Bldg. 74, P.O. Box 5000, Upton, NY 11973-5000*

[6] *Department of Physics and Astronomy, Rutgers, The State University of New Jersey, 136 Frelinghuysen Road, Piscataway, New Jersey, 08854, USA*

[7] *Center for High Pressure Science and Technology Advanced Research (HPSTAR), Beijing 100094, China*

Thermal ammonolysis of quasi-two-dimensional (quasi-2D) CoTa$_2$O$_6$ yields the O$^{2-}$/N$^{3-}$ and anionic vacancy ordered Co$^{2+}$Ta$^{5+}_2$O$_{6-x}$N$_{2x/3}$□$_{x/3}$ ($x \leq 0.15$) that exhibits a transition from antiferromagnetism to defect engineered above room-temperature ferromagnetism as evidenced by diffraction, spectroscopic and magnetic characterizations. First-principles calculations reveal the origin of ferromagnetism is a particular CoO$_5$N configuration with N located at Wyckoff position 8*j*, which breaks mirror symmetry about ab plane. A pressure-induced electronic phase transition is also predicted at around 24.5 GPa, accompanied by insulator-to-metal transition and magnetic moment vanishing.

## I. INTRODUCTION

Room-temperature (RT) ferromagnetic (FM) insulators have opened a fascinating field of multiferroics and spintronic devices [1-5]. In the recent past, diluted magnetic semiconductors (DMS) played a major role in this field. Magnetic ions doped nonmagnetic oxides, such as Co-doped TiO$_2$, Cr, Mn, Ni, or Nd-doped ZnO [6-10], can induce long-range FM order deriving from the strong electron exchange interactions of the guest dopant. Later on, it was found that vacancy-driven defect engineering, such as the incorporation of oxygen or transition metal vacancies, can also facilitate RT FM interactions in nonmagnetic oxides such as CaO, MgO, ZnO, TiO$_2$, Al$_2$O$_3$, and others [11-15]. Recently, the observation of emergent long-range FM order in insulating van der Waals (vdW) layered materials including CrI$_3$, Cr$_2$Ge$_2$Te$_6$, Fe$_3$GeTe$_2$, and Cr$_2$Te$_3$ has inspired new research on two-dimensional (2D) FM materials [16-19]. For examples, exfoliated monolayer CrI$_3$ displays intrinsic 2D ferromagnetism with Curie temperature ($T_C$) of 45 K [16]; the 2D Heisenberg ferromagnet Cr$_2$Ge$_2$Te$_6$ shows stable magnetic moments at small magnetic field of 0.075 T and strong dimensionality effect [17]. Thin Fe$_3$GeTe$_2$ undergoes a structure transformation from 3D to 2D Ising ferromagnetism, followed by a decline of the $T_C$ from 207 to 130 K in the monolayer [18]. Another breakthrough in six unit-cell thick (~ 7.1 nm) Cr$_2$Te$_3$ realizes near RT ferromagnetism ($T_C$ = 280 K), suggesting thickness-dependent $T_C$ modulation in typical vdW crystals [19]. 2D transition metal dichalcogenides (TMDs) materials have also spawned remarkable interests due to their robust RT ferromagnetism, such as VSe$_2$, MoS$_2$, MoTe$_2$, and WS$_2$ [20-22]. Above RT ferromagnetism ($T_C$ up to 470 K) has been found in monolayer 1T-VSe$_2$ (trigonal lattice and octahedral coordination) originated from Se vacancies [20]. MoS$_2$ and WS$_2$ nanosheets exhibit high $T_C$ ferromagnetism (865 K for MoS$_2$ and 820 K for WS$_2$) stimulated by defects and dislocations engineered during their growth process [21], while the ferromagnetism ($T_C$ above 400 K) of MoTe$_2$ nanosheets is achieved by anionic defects via synergetic control of Ta doping, thickness, and lateral size [22].

Furthermore, hydrostatic pressure offers a systematic control for tuning electrical and magnetic properties of 2D vdW and TMDs materials. The magnetic response of 2D chalcogenides can be tuned by external pressure as discovered in the confined two-layered CrI$_3$, which revealed a pressure-induced antiferromagnetic (AFM) to FM transition at 2.7 GPa, providing thermodynamic freedom for optimized performance [23-25]. Multilayered MoS$_2$ and WS$_2$ undergo a semiconductor-to-metal transition at ~19 GPa and ~22 GPa, respectively [26,27], originating from sulfur-sulfur interactions as the interlayer spacing reduces with the increment of pressure.

These findings point out an effective approach to search for RT FM insulators, semiconductors or metals in 2D materials by dimensional controlling, defect engineering, and/or compressing. In contrast to the flourishing success in low-dimensional chalcogenides [4,28-31], there are relatively few studies devoted to FM insulator design in the chemically benign (stable and easy synthesis) 2D oxides [32]. In this work, we ammoniated the trirutile-type CoTa$_2$O$_6$ (CTO) and successfully realized the transition from initial antiferromagnetism to above RT ferromagnetism via anionic defect engineering in the nitridated product. The crystal and electronic structures of as-made samples were extensively studied by diffraction and spectroscopic techniques. Defect-dependent magnetism evolution was also experimentally and theoretically studied. First-principles calculations were conducted to reveal the origin of the above RT magnetism, estimate the band gap, and evaluate the evolution of magnetism and conductivity under high pressure.

*Corresponding author: zengyj@hdu.edu.cn
†yaodaox@mail.sysu.edu.cn
‡limanrong@mail.sysu.edu.cn

## II. METHODS

### A. Synthesis

The CTO precursor was synthesized by solid-state reactions according to the previous report [33]. Stoichiometric mixture of the raw materials CoO (99.99%, Aladdin) and $Ta_2O_5$ (99.99%, Aladdin) was ground in an agate mortar before shifted to an alumina crucible. The sample was heated to 1000°C (heating rate of 5°C min$^{-1}$), followed by increasing the temperature to 1050°C (heating rate of 1°C min$^{-1}$) and held for 12 h before programmable cooling (cooling rate of 5°C min$^{-1}$) to RT. The obtained powder was placed in quartz crucibles and heated in a tube furnace under an ammonia flow of 150 mL min$^{-1}$ for 12 h at 650, 700, 750, 800, 850, and 900°C, respectively, with both heating and cooling rate of 5°C min$^{-1}$. The as-nitrided $CoTa_2O_{6-x}N_{2x/3}$ powders are hereafter denoted as CTON-650, CTON-700, CTON-750, CTON-800, CTON-850, and CTON-900, respectively, according to the reaction temperature.

### B. Chemical and Spectroscopic Characterizations

X-ray powder diffraction (XRD) data were collected at RT on a Panalytical Empyrean diffractometer with Cu $K\alpha$ radiation ($\lambda$ = 1.54187 Å) operated at 40 kV and 40 kA, in the $2\theta$ ranging from 10° to 120° with a step size of 0.02° and scanning time of 3.93 s per step. The TOPAS-Academic V6 software package was applied to perform diffraction data analysis and Rietveld refinement [34]. Thermogravimetric analysis (TGA) was sequentially performed with an SDT Q600 instrument in the temperature range of 30-900°C at a speed of 10°C min$^{-1}$ in air. Transmission electron microscopy and energy-dispersive X-ray spectroscopy (TEM-EDS) spectra/mapping were obtained from a JEOL JEM-ARM200F with acceleration voltages of 200 kV. UV-visible diffuse reflectance spectra (UV-Vis DRS) were collected on the Varian Cary 5000 UV-vis-NIR spectrometer in the visible range of 200-800 nm, using $BaSO_4$ as the reference of 100% reflectance. X-ray absorption near edge spectroscopy (XANES) measurements on CTO and CTON-800 were performed at the Brookhaven NSLS-II on beamline 7BM QAS using both transmission and fluorescent modes with a simultaneous standard for precision energy calibration. Standard linear background and post-edge normalization to unity were used in the analysis. Many of the standard spectra were collected on NSLS-I on beamline X19A. X-ray photoelectron spectroscopy (XPS) was conducted on a Nexsa XPS system equipped with a monochromatic Al $K\alpha$ X-ray source ($hv$ = 1486.6 eV). All binding energies were calibrated using surface contaminant carbon (C 1$s$ = 284.8 eV) as a standard. Fourier transform infrared spectroscopy (FT-IR) were recorded in the wavenumber range of 4000-400 cm$^{-1}$ with a SHIMADZU IRAFFINITY-1S spectrometer at RT, for which the samples were mixed with KBr (a mass ratio of 1 : 200) and pressed into a pellet.

### C. Magnetic Measurements

Magnetic measurements were implemented with a Physical Properties Measurement System (PPMS, Quantum Design). The magnetic susceptibility ($\chi$) was measured in zero field cooled (ZFC) and field cooled (FC) modes in a temperature scope of 5-300 K under 0.1 T magnetic field ($H$). Field dependent isothermal magnetization was measured between 10 and 550 K under an applied magnetic field from -4 to 4 T.

### D. Computational Methods

The first-principles calculations were performed using the VASP code [35], with the exchange-correlation interaction suggested by Perdew, Burke and Ernzerhof (PBE) [36], and pseudopotentials constructed by projector augmented-wave (PAW) method [37]. The cutoff energy for plane wave expansion was 700 eV. A 15 × 15 × 9 (9 × 5 × 5) Monkhorst-Pack grid was used for integration in the Brillouin zone (BZ) for the bulk (supercell) [38]. To account for the combined effects of nitrogen dopant and oxygen vacancy, a 1 × 5 × 1 supercell was constructed, with two oxygen atoms replaced by nitrogen atoms, leaving one oxygen vacancy and a doping ratio being about $x = 0.3$ in $CoTa_2O_{6-x}N_{2x/3}\square_{x/3}$, where $\square$ represents the oxygen vacancy and is introduced in light of charge balance. There are $C_{60}^3$ ways of arranging the two nitrogen atoms and one oxygen vacancy in the supercell, and here we considered only two cases: oxygen vacancy located at Wyckoff position 4$f$ and 8$j$, respectively. One nitrogen atom is near to the oxygen vacancy, while the other is away from it. The atoms were allowed to relax until the forces were smaller than 0.01 eV/Å. The tolerance of energy for electronic relaxation was $1.0 \times 10^{-9}$ eV. The LDA+U method (LDAU) introduced by Dudarev *et al.* [39] was used to account for the on-site Coulomb interaction of Co, with $U = 0$, 2.5 and 4 eV, respectively.

## III. RESULTS AND DISCUSSION

### A. Phase and structural characterizations

Figure 1(a) (bottom) shows XRD patterns of CTO, which is consistent with the reported tetragonal $P4_2/mnm$ cell [40]. The precursor was treated under $NH_3$ flow at temperatures range from 650 to 900°C. Attempts to azotize CTO below 650°C or above 800°C were unsuccessful, since the precursor remained either unchanged (< 650°C) or started to decompose (> 800°C). Figure 1(a) displays the XRD patterns of CTO after azotized at different temperatures. The normalized XRD patterns of CTO and CTON-800 are compared in Fig. S1 (Supporting Information online)**,** which indicates slightly weakened intensity of most XRD peaks in CTON-800 except the one at 21.1°. In addition, the color of $CoTa_2O_{6-x}N_{2x/3}$ successively gets darker with temperature increment (Fig. 1(a)), starting from pink for CTO, passing through dark pink (CTON-650 and CTON-700) to grey (CTON-750 and CTON-800). This variation of optical absorption indicates the enhanced $N^{3-}$ incorporation into CTO at elevated temperatures [41,42], as further corroborated by TGA in air (Fig. S2) discussed below.

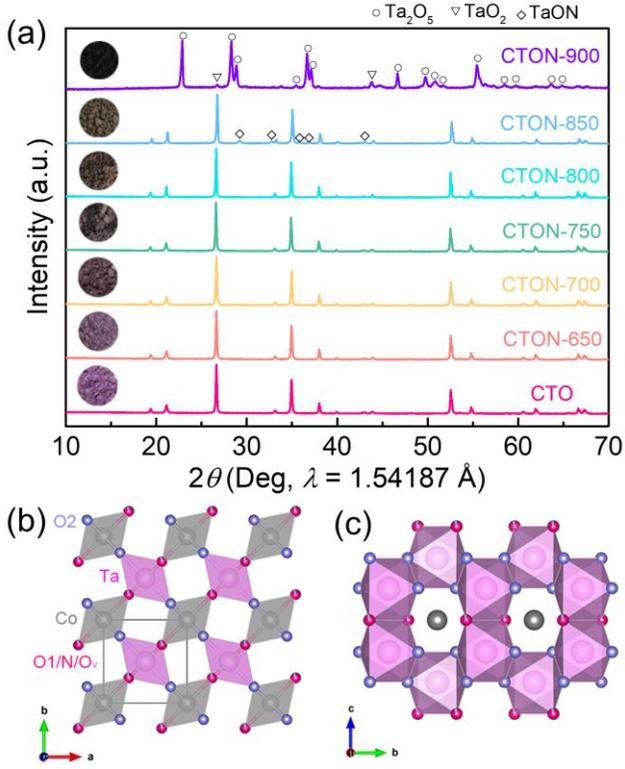

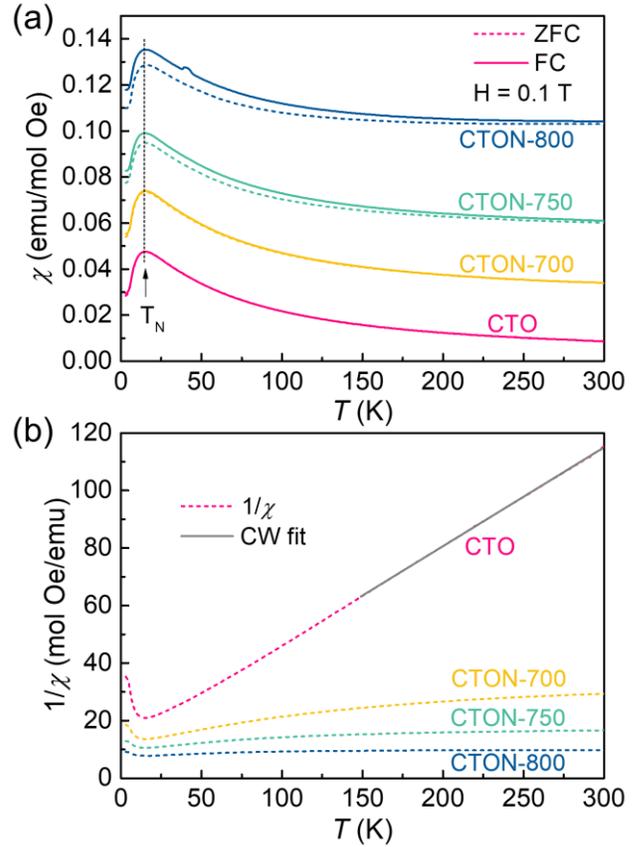

FIG. 1. (a) XRD patterns of $CoTa_2O_{6-x}N_{2x/3}\square_{x/3}$ prepared at various temperatures under $NH_3$ flow. The XRD patterns of CTO precursor is put on the bottom for comparison. Left insert shows the color evolution of $CoTa_2O_{6-x}N_{2x/3}\square_{x/3}$ powder. Crystal structure of $CoTa_2O_{6-x}N_{2x/3}\square_{x/3}$, where the view of (b) layered structure along $c$-axis and (c) honeycomb-like ring along $a$-axis. The light silver octahedra are $CoO_{6-x}N_{2x/3}$, the light pink octahedra are $TaO_{6-x}N_{2x/3}$, Co is represented by silvery gray, Ta-light pink, O1-rosy red, N/□ (oxygen vacancies)-cyan, O2-violet. The N atoms and oxygen vacancies are distributed at the O1 site.

FIG. 2. Magnetic properties of $CoTa_2O_{6-x}N_x\square_{x/3}$. (a) Temperature dependence of ZFC (short dash line) and FC (solid line) $\chi$ curves of CTO, CTON-700, CTON-750 and CTON-800, measured at 0.1 T. (b) The ZFC $\chi^{-1}$ vs T curves of CTO, CTON-700, CTON-750 and CTON-800, measured at 0.1 T. The solid line shows the CW fitting of paramagnetic region between 150 and 300 K.

The N incorporation in $CoTa_2O_{6-x}N_{2x/3}$ was evaluated by TGA as depicted in Fig. S2. The weight variation below 200°C in Fig. S2(a-c) is attributed to physical absorption. All oxynitrides undergo a stage of mass gain below ~ 700°C, owing to the release of nitrogen and re-oxidation process. (Fig. S2(b-d)) [43]. The TGA curves exhibit further weight loss after the oxidation process above 700°C, probably due to the decomposition of intermediate compounds [44]. The nitrogen-richest compound CTON-800 is more reactive, as evidenced by the onset of oxidation-induced weight gaining started above 320°C compared with that from 560 and 446°C for CTON-700 and CTON-750, respectively. Accordingly, the $x$ value in $CoTa_2O_{6-x}N_{2x/3}$ can be estimated from these data as each weight increment is equal to the difference between released $N_2$ and up-take $O_2$ [45,46]. Figure S3 shows XRD patterns for $CoTa_2O_{6-x}N_{2x/3}$ samples after TGA measurements, indicating that the nitrided samples reverted back to CTO. There is no evidence of $Co^{3+}$ formation during the oxygenating process. The TGA weight gains were determined to be 0.0174(3)%, 0.0627(12)%, and 0.1859(7)% corresponding to CTON-700, CTON-750 and CTON-800 with $x = 0.015(1), 0.05(1)$ and $0.15(1)$, respectively. To maintain charge balance in $CoTa_2O_{6-x}N_{2x/3}$, the substitution of $O^{2-}$ by $N^{3-}$ simultaneously generates anion vacancies since $Co^{3+}$ is unlikely to survive in the reducing $NH_3$ atmosphere, giving nominal formula of $CoTa_2O_{6-x}N_{2x/3}\square_{x/3}$ (where □ denotes the anionic vacancy). TEM-EDS analysis confirms the Co/Ta ratio (~ 1/1.92(6)) of CTON-800 and homogeneous Co and Ta distribution (Fig. S4). As N content increase, the band gap ($E_g$) exhibited a narrowing from the initial 3.22 eV to 3.02 eV, which is estimated by UV-Vis DRS (Fig. S5).

Figure S6(a) shows the refined plots of XRD patterns for CTO in the reported $P4_2/mnm$ space group (No. 136, $a = 4.7375$ (4) Å, $c = 9.1764$ (9) Å, $V = 205.95$ (5) Å$^3$, $Z = 2$) [33,47]. $CoTa_2O_{6-x}N_{2x/3}\square_{x/3}$ ($x = 0.015, 0.05$ and $0.15$) are isostructural to CTO, and the same starting model was applied for their structure refinements as exhibited in Fig. S6(b-d). The refined crystallographic parameters and selected interatomic distances and angles are summarized in Tables S1 and S2. Although the effective ionic radius of $N^{3-}$ ion ($r_{eff} = 1.46$ Å) is slightly larger than that of $O^{2-}$ ($r_{eff} = 1.38$ Å) [48], the lattice parameter tends to be constant in different N content, resulting from the associated increase of anionic vacancies as shown in Fig. S7. Previous studies show that high-valency metal cation with smaller ionic radius can sometimes form to keep charge balance in a number of perovskite oxynitrides [49-51]. However, the ammonolysis process drives the more electronegative 3$d$ transition metals Fe, Co, Ni and Cu to form interstitial nitrides with low positive oxidation states [50]. It is thus unlikely to form $Co^{3+}$ by ammonolysis of $Co^{2+}Ta^{5+}_2O_6$. The formal oxidation state of cations in $CoTa_2O_{6-x}N_{2x/3}\square_{x/3}$ is further supported by XANES (Fig. S8) and XPS (Fig. S9) measurements. FT-IR results also confirm the interactions among chemical bonds in $CoTa_2O_{6-x}N_{2x/3}\square_{x/3}$ (Fig. S10). In addition, the 2D honeycomb-layered $CoTa_2O_{6-x}N_{2x/3}\square_{x/3}$ compounds have the same six-coordinated Co and Ta ions (Fig. 1(b, c)) as in CTO. The O1/N anions and vacancies were distributed at the $4f$ $(x, y, 0)$ sites, while O2 ions fully occupy $8j$ $(x, y, z)$ sites during refinements according to the calculated results (vide infra).

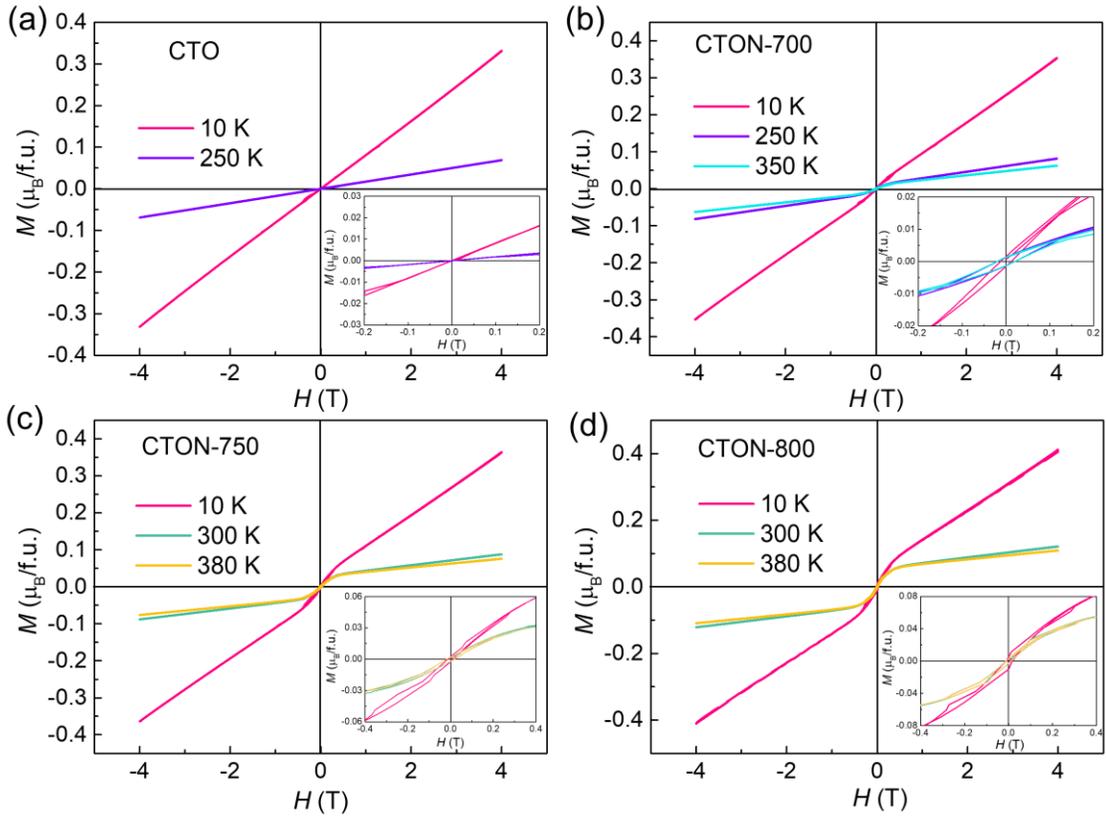

FIG. 3. Isothermal magnetization curves of $CoTa_2O_{6-x}N_x\square_{x/3}$ at 10, 250, 300, 350 and 380 K between -4 and 4 T. (a) CTO; (b) CTON-700; (c) CTON-750; (d) CTON-800. The insets in (a) and (b) show the enlarged area between -0.2 and 0.2 T, in (c) and (d) show between -0.4 and 0.4 T.

**B. Magnetic properties**

Figure 2(a) shows the temperature-dependent magnetic susceptibility $\chi(T)$ of $CoTa_2O_{6-x}N_{2x/3}\square_{x/3}$ ($x$ = 0.015, 0.05 and 0.15) with $T$ = 5-300 K measured at $H$ = 0.1 T in ZFC and FC modes. It is clearly seen that bulge peaks of all samples arise in the ZFC and FC curves (Fig. 2(a)), illustrating the predominance of AFM phase transition at $T_N$ = 14.7 K, which is close to the literature result [52]. Notably, unlike CTO, the ZFC and FC curves of N-contained samples do not overlap until up to 300 K (Fig. 2(a)) in analogous to ZnO nanosheets [53], indicating the existence of RT ferromagnetism. Above 150 K, the ZFC $1/\chi - T$ curve for CTO can be fitted with the Curie-Weiss (CW) law, $1/\chi = (T - \theta_W)/C$, where $C = \mu_{eff}^2/8$ is the Curie constant, $\theta_W$ is Weiss temperature, and $\mu_{eff}$ is the effective magnetic moment (Fig. 2(b)). The $\theta_W$ = -35 K is consistent with AFM order. The values of $\mu_{eff}$ for CTO was calculated to be 4.82 $\mu_B$, which agrees well to the typical value for high-spin $Co^{2+}$ ($S$ = 3/2) in octahedral geometry. In Fig. 2(b), in contrast to CTO, the $1/\chi - T$ curves for CTON-700, CTON-750 and CTON-800 exhibit nearly temperature independent behavior at $T > 50$ K and do not obey the CW law, implying different magnetic interaction mechanism in anionic vacancy rich cases.

Additional characterization of the nature of magnetic transition in $CoTa_2O_{6-x}N_{2x/3}\square_{x/3}$ samples was achieved through isothermal field-dependent magnetization $M(H)$ measured between 5 and 550 K and applied magnetic field from -4 to 4 T (Fig. 3 and Fig. S11). As shown in Fig. 3(a), the magnetization curves for CTO show no hysteresis at 10 and 250 K, supported the AFM interactions. Interestingly, CTON-700, CTON-750, and CTON-800 samples show slight hysteretic loops over the experimental temperature range, indicating RT FM feature (Fig. 3(b-d)). Their saturation magnetizations slightly decrease with increasing temperature, which is attributed to the changes in the magnetic moment of cobalt beside anion vacancies. Note that the nitrided series have larger slope values ($M/H$) in $0 < H < 0.5$ T than those in $H > 0.5$ T. We assume that the AFM coupling is weakened with enhanced nitrogen incorporation, and it might be accompanied by the appearance of weak FM states in $0 < H < 0.5$ T as displayed in Fig. S12. For DMS materials, the formation of oxygen vacancies lowers the valence of the metal ions and thus entangles ferromagnetism. While the vdW materials such as $CrI_3$ and $VSe_2$ conceive intrinsic ferromagnetism in exfoliated monolayer [16,20]. As for $CoTa_2O_{6-x}N_{2x/3}\square_{x/3}$, neither valence transition nor thickness-dependent layer interface exists, the emergent RT ferromagnetism differs from known studies. Therefore, theoretical calculations were performed to figure out the origin of emerging ferromagnetism.

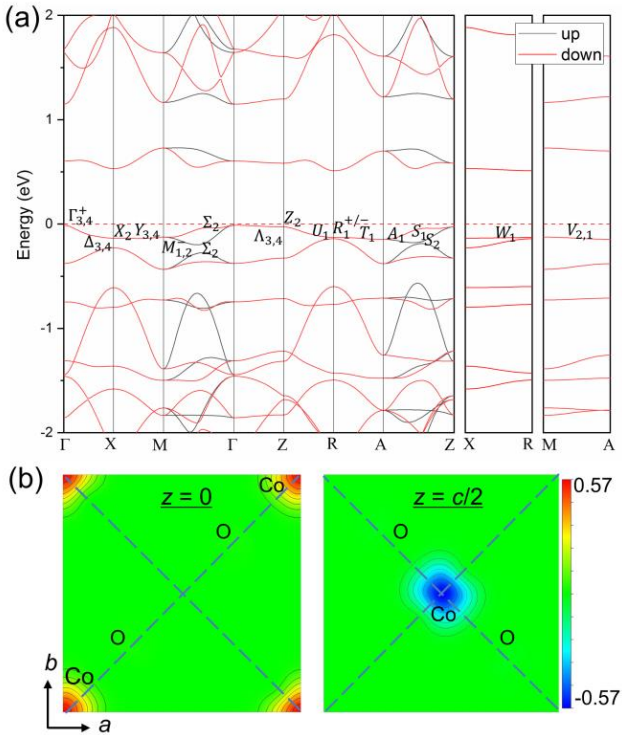

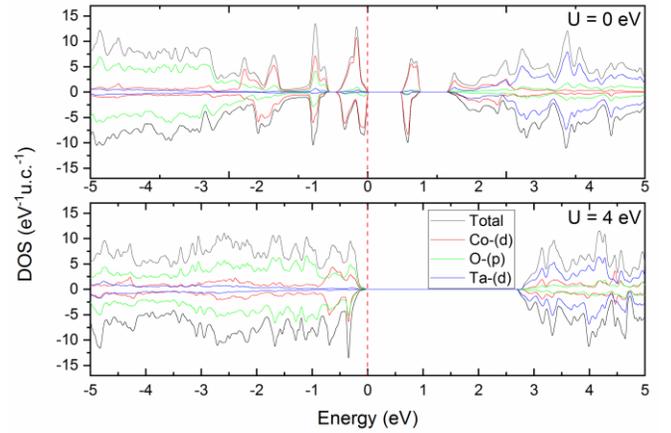

FIG. 4. (a) The calculated spin polarized band structure, without considering the on-site Coulomb interaction of Co atoms ($U = 0$ eV). The horizontal dashed line denotes the Fermi energy. The representation labels are the same as those in Ref. [57]. The representation of R can be either $R_1^+$ or $R_1^-$. (b) The magnetization density of CTO on the (001) plane at $z = 0$ and $c/2$, respectively, $c$ is the lattice constant along $z$ direction. The positions of Co and O atoms are indicated.

### C. Theoretical calculations

#### 1. Bulk electronic structure: k-dependent spin splitting

Although the magnetic structure of CTO is controversial and not fully determined [54], it was corroborated that the magnetic moments of Co atoms lie in the $ab$ plane and antiferromagnetically coupled along [110] on $z = 0$ and [-110] on $z = c/2$, with propagation vectors ($\pm 1/4$, $1/4$, $1/4$). To model such a magnetic structure accurately would require a supercell containing 576 atoms, which is beyond the capability of modern first-principles calculation method. Thus, in our calculations only spin polarization calculation is adopted, and some new and meaningful results are obtained.

The band structure of AFM CTO shows $k$-dependent spin splitting [55], which is different from those of conventional antiferromagnets whose spin-up and spin-down energy levels are degenerate along all $k$-points in the BZ. Here, the spin-up and spin-down energy levels of CTO are nondegenerate along

FIG. 5. The density of states near Fermi energy of CTO, without (upper panel, $U = 0$ eV) and with (lower panel, $U = 4.0$ eV) on-site Coulomb interaction of Co atoms. The vertical dashed line denotes the Fermi energy.

$\Gamma M$ ($u$, $u$, 0) and ZA ($v$, $v$, 0.5) (where $0 < u$, $v < 0.5$), both of which are along the (110) direction of the BZ. This spin splitting energy is as large as 150 meV for the valence band below Fermi energy, much larger than the splitting caused by spin orbit coupling for $3d$ transition metals [56]. This behavior can be understood as follows [57]. In AFM spin polarized calculations the magnetic moments of Co atoms are directed along $z$ direction, which reduces the Shubnikov group of CTO from paramagnetic $P4_2/mnm1'$ (BNS [58]: 136.496) to AFM $P4_2'/mnm'$ (BNS: 136.499). In the former case, time reversal symmetry ensures double degeneracy for all energy levels at all $k$ points, namely spin-up and spin-down energy levels are degenerate. In the latter case, however, time reversal symmetry is broken but its combination of some space group symmetry operations exists. In particular, the group of wave vector at $\Sigma$ ($\alpha$, $\alpha$, 0) ($0 < \alpha < \pi/a$), for example, is $\{(1|0$ , 0, 0), (2 - $xy$|0, 0, 0)', (m$z$| 0, 0, 0),(m$xy$|0, 0, 0)'$\} = C_{2v}$ ($C_s$), of which the single-valued representations ($\Sigma_1$, $\Sigma_2$) are both one-dimensional [57]. Thus the energy levels at $\Sigma$ are nondegenerate, as revealed by the spin-splitting of the energy bands along $\Gamma M$. Similar explanations can be applied to other $k$ vectors and the detailed labelling of the representations are shown in Fig. 4(a).

Figure 4(b) provides the calculated magnetization density around the two Co atoms. The magnetization is defined as $m(r) = \rho^\uparrow(r) - \rho^\downarrow(r)$, where $\rho^\uparrow(r)$ and $\rho^\downarrow(r)$ are spin-up and spin-down electron densities, respectively. The magnetization shows clear anisotropy in the $ab$ plane, indicating the magnetic moments of Co atoms at the corner tend to align towards [110] or [-1-10] and those at the center towards [1-10] or [-110], but the absolute direction cannot be determined.

The $k$-dependent spin-splitting causes the density of states (DOS) to be asymmetric for spin-up and spin-down channels, as shown in Fig. 5. Here both the PBE and LDAU results are shown. The PBE result underestimates the band gap to about 0.5 eV, which is a well-known problem of density functional theory. The LDA+U method remedies this problem effectively and gives a band gap about 2.7 eV, which is close to the experimental value, too insulating to be measured within the ammeter range. However, the crystal field splitting of the Co $d$ orbitals, caused by the oxygen octahedron, appears clearly near the Fermi energy in the PBE derived DOS, but is lost in the LDAU one due to the large on-site Coulomb interaction. Since the magnetic susceptibility and crystal field splitting are ground state properties, while band gap is related with excitation states. Moreover, density functional theory [59,60] is expected to work well only for ground state property, we propose to rely on the PBE results to discuss the magnetic property.

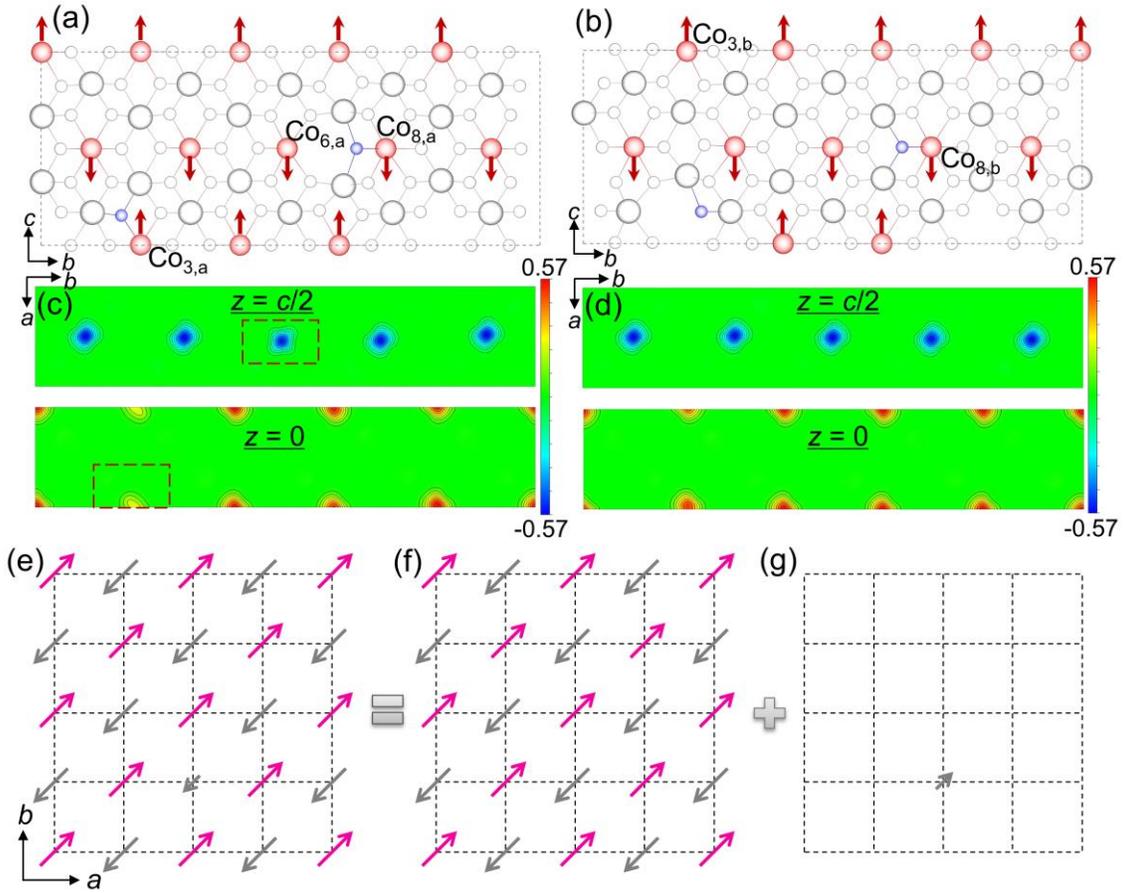

FIG. 6. The supercell with two nitrogen dopant and one oxygen vacancy, corresponding to $CoTa_2O_{6-x}N_{2x/3}\square_{x/3}$, $x = 0.3$. The oxygen vacancy can be located at (a) Wyckoff position $4f$ ($x$, $x$, 0) or (b) $8j$ ($x$, $x$, z). The red, grey, white and blue spheres represent cobalt, tantalum, oxygen and nitrogen atoms, respectively. The arrows on cobalt atoms indicate the magnetic moments. The corresponding magnetization density on the (001) plane, with oxygen vacancy at (c) $4f$ and (d) $8j$, respectively. The $z = c/2$ case is in the upper panel and $z = 0$ case in the lower panel. Illustration of the fact that a (e) ferrimagnet can be decomposed into an (f) antiferromagnet and a (g) ferromagnet.

### 2. The effect of nitrogen dopant on the electronic structure: from antiferromagnet to ferrimagnet

The $x = 0.3$ case of $CoTa_2O_{6-x}N_{2x/3}\square_{x/3}$ other than the experimental $x = 0.15$ is considered in our calculations to avoid too large supercell, with the crystal structure shown in Fig. 6(a, b). This increase of dopant ratio should not cause trouble because the distance between vacancies is already small in $x = 0.3$ case. Since there are two types of oxygen atoms in the bulk crystal of CTO, with location at Wyckoff position $4f$ or $8j$, the oxygen vacancy is thus put to either $4f$ or $8j$. One nitrogen atom is near to the oxygen vacancy, at the same Wyckoff position, and the other nitrogen is away from the oxygen vacancy, at the opposite Wyckoff position.

Our first-principles calculations reveal that when the oxygen vacancy is located at the $4f$ sites, there is a net magnetic moment about -1.8 $\mu_B$ in the supercell, corresponding to a saturation magnetization of 0.18 $\mu_B$ per Co atom. However, in the $8j$ oxygen vacancy case, there is no net magnetic moment in the supercell. Apparently, anions and vacancies are ordered in $CoTa_2O_{6-x}N_{2x/3}\square_{x/3}$ as reasonably restrained during the structural refinements in Fig. S6. We show the magnetization densities in the $ab$ plane in Fig. 6(c, d), since the magnetic moments of Co atoms mainly interact and lie in this plane. From Fig. 6(c) it is clear that the magnetic moment of one Co atom with $CoO_5N$ coordination ($Co_{3,a}$) is reduced in the $4f$ case. We further find that the magnetic moments of other two Co atoms ($Co_{8,a}$ and $Co_{8,b}$) with $CoO_5N$ coordination remain the same as that of the bulk Co atom. This different behavior comes from the fact, that in $Co_{3,a}$, the Co-N bond is not located in $ab$ plane, and about which the mirror symmetry is broken. The shape of magnetization density of $Co_{6,a}$ with $CoO_5$ coordination changes a little but is still along [1-10] or [-110], and the magnitude of magnetic moment is not changed. There is no visible change of magnetization density of $Co_{3,b}$ with $CoO_4N$ coordination, and the magnitude of magnetic moment remains the same as those in the bulk. Physically this behavior can be explained by inspecting the partial density of state (PDOS) of $d$ orbitals of the Co atom, which reflects the effect of the crystal field of nearby coordination ($O_6$, $O_5N$, $O_4N$, *etc.*), as exemplified in Fig.

S13.

We are now safe to arrive at the conclusion that the inclusion of nitrogen dopant in CTO, with Co-N bond not in *ab* plane and coordination of Co atom with $O_5N$, can turn the sample from antiferromagnet to ferrimagnet, and the slight hysteretic loop $M(H)$ in Fig. 3 is a direct consequence of this transition. As shown in Fig. 6(e-g), a ferrimagnet can be deconstructed into an antiferromagnet and a ferromagnet, and the magnetization curve under an external field can be decomposed into an AFM component $M_{AFM}$ and a FM component $M_{FM}$. Note that in the FM sublattice, the number of magnetic moments per unit volume $n_{FM}$ is much smaller than the number of magnetic moments per unit volume in the AFM sublattice $n_{AFM}$, and is dependent on the doping ratio $x$. Since there is no experimental evidence that the $CoO_5N$ is ordered, and considering the low density of "magnetic ion" $CoO_5N$, the ferrimagnet model adopted by calculation is not able to explain the high Curie temperature and ferromagnetism found here. However, the model indicates that there might be some mechanism similar to that in DMS, the only difference being that in DMS the host is nonmagnetic semiconductor [61], but here the host is an antiferromagnet. This particular DMS system with an AFM host deserves further study. Note that the saturation magnetization is proportional to the magnetic moments per unit volume, thus the dilute magnetic ions should reach saturation at much lower external field than that of the AFM host. The discontinuity at 0.5 T in the magnetization curves is thus understood as the saturation field of the dilute magnetic ions part.

*3. Predicted pressure effect on the electronic structure*

Extrinsic pressure has proven to be powerful to tune the properties in 2D materials [23-25]. Here we further analyzed the behavior of electronic structure of CTO under high pressure, assuming there is no structural phase transition, that is, the space group $P4_2/mnm$ is maintained. Figure 7(a) shows the variation of total energy with respect to unit cell volume, which has an inflection point near $V = 190$ Å$^3$, indicating the existence of electronic phase transition. By fitting the data to Murnaghan equation for $V < 190$ Å$^3$ (phase 2) and $V > 190$ Å$^3$ (phase 1), respectively, we obtained the pressure dependence of the relative enthalpies ($\Delta H$, or Gibbs free energies at zero temperature) between the two phases (Fig. S14). The phase transition occurs at pressure about 24.5 GPa. The variation of volume with pressure is shown in Fig. S15, from which the discontinuity of volume at the transition point is clearly seen, and the pressure at a certain volume can be determined. The phase transition is also manifested in the variation of lattice constants with respect to volume (or pressure), as shown in Fig. S16.

CTO becomes metallic in phase 2, as shown in Fig. S17, indicating that the phase transition is insulator-to-metal transition. The most striking feature here is that after entering into the metallic state, the magnetic moment of Co disappears, and CTO becomes nonmagnetic. This is manifested by the drop to zero of the values of the magnetic moment of Co in Fig. 7(a), and the disappearance of the $k$-dependent splitting between spin-up and spin-down energy bands along ΓM and ZA. Here we note that before entering into the nonmagnetic metallic phase, CTO is in a border phase with a low spin state for $Co^{2+}$, as manifested by the four data points near $V = 190$ Å$^3$ in the magnetic moment-volume curve in Fig. 7(a). This transition is due to that the crystal field splitting between $e_g$ and $t_{2g}$ orbitals (actually $e_g$ and $t_{2g}$ are nondegenerate here, due to the $D_{2h}$ site symmetry of Co, but the deviation may not be large) becomes larger than the Coulomb interaction, as pressure increases. Note that for $Co^{2+}$ with six-coordination, the radius for high-spin and low-spin states are 0.745 and 0.65 Å, respectively [62]. The radius of three-coordinated $O^{2-}$ is 1.36 Å. We thus have bond length of Co-O bond to be 2.105 and 2.01 Å, respectively. The variation of calculated Co-O bond lengths with respect to volume is shown in Fig. 7(b). It can be seen that the bond length of Co-O bond with low-spin $Co^{2+}$ state just locates at the position where CTO enters in the border phase. The reason why as pressure increases (volume decreases) the magnetic moment of Co becomes zero is still not clear. However, the PDOS of Co in such high pressure condition shows that the PDOS of $d_{x2-y2}$ orbital is zero, indicating that this orbital might not be occupied. Further experimental work is needed to explain this phenomenon. The behavior of electronic structure of nitrogen doped CTO under high pressure is out of the scope of this study, it's anticipated that similar insulator-to-metal transition can occur. However, due to the presence of nitrogen dopant and oxygen vacancy, the magnetic moment of Co surrounded by those vacancies may not vanish, and the sample would turn into a FM metal.

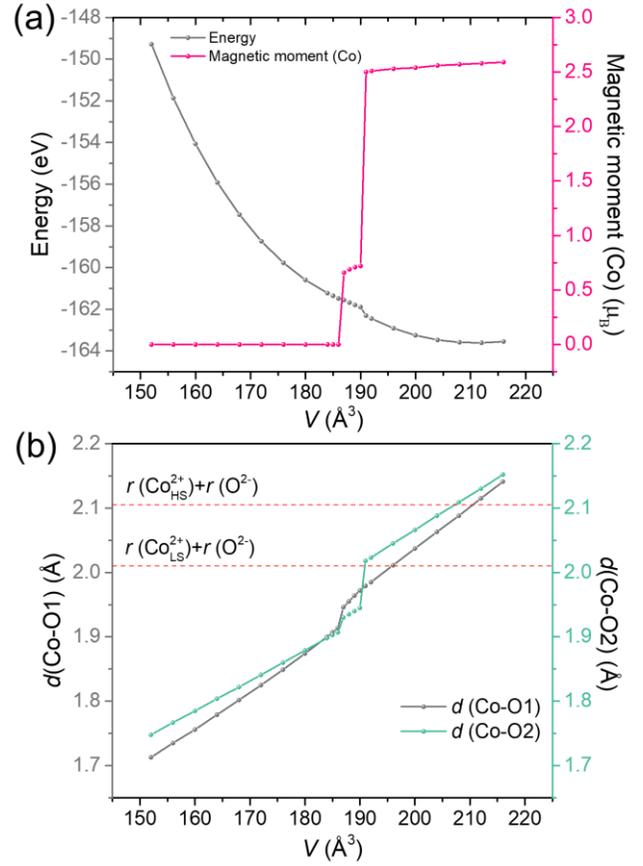

FIG. 7. (a) The variation of total energy and the average absolute value of magnetic moments of Co atoms in CTO with the unit cell volume. Near $V = 190$ Å$^3$ there is a clear inflection point in the E-V curve, associated with discontinuous change of magnetic moment of Co atom. (b) The evolution of the bond length of Co-O1 bond (in *ab* plane) and Co-O2 bond (not in *ab* plane) with unit cell volume. The sum of the radius of $Co^{2+}$ with coordination six and $O^{2-}$ with coordination three are shown, for high-spin and low-spin Co states, respectively.

## IV. SUMMARY

In summary, we have successfully synthesized the anionic ($O^{2-}/N^{3-}$) and vacancy ordered oxynitrides $CoTa_2O_{6-x}N_{2x/3}\square_{x/3}$ ($x$ = 0.015, 0.05 and 0.15) via thermal ammonolysis of $CoTa_2O_6$, in which the nitrogen contents were adjusted by reaction temperature. The nitrided samples demonstrate narrowing band gaps, from 3.22 eV for $CoTa_2O_6$ to 3.02 eV for $CoTa_2O_{5.85}N_{0.1}\square_{0.15}$, and above room-temperature ferromagnetic response. First-principles calculations reveal that the emerging ferromagnetism is engineered by the $CoO_5N$ configuration caused by nitrogen doping, which evoke the transformation of octahedral configuration of cobalt and energy level, leading to change in magnetic moment. An insulator-to-metal transition is also predicted to arise around 24.5 GPa due to electronic phase transition accompanied with the collapse of magnetic moments. These discoveries identified a simple and effective approach to realize practical ferromagnetism, and are expected to forward a new mechanism of magnetic transitions, and advance exploration of novel 2D oxides for spintronic devices.

## ACKNOWLEDGMENTS

This work was supported by the National Science Foundation of China (NSFC-22090041 and 21875287), the Program for Guangdong Introducing Innovative and Entrepreneurial Teams (2017ZT07C069). Y. Z. and D. X. Y. are supported by NKRDPC-2018YFA0306001, NKRDPC-2017YFA0206203, NSFC-11974432, NSFG-2019A1515011337, and Leading Talent Program of Guangdong Special Projects, and National Science Foundation (NSF-DMR-1507252). Work at Brookhaven National Laboratory was supported by the DOEBES (DE-SC0012704). The theoretical part in this work was performed on TianHe-2 at the National Supercomputer Center in Guangzhou. The authors would like to thank Dr. Z. Deng at Institute of Physics, Chinese Academy of Sciences for helpful discussion of the magnetic measurements; Mr. K. Huang at School of Physical and Mathematical Sciences, Nanyang Technological University for valuable discussions.

# Supporting Information

# Defect Engineered Room-Temperature Ferromagnetism in Quasi-Two-Dimensional Nitrided $CoTa_2O_6$

Yalin Ma,[1] Shuang Zhao,[1] Xiao Zhou,[1] Yijie Zeng,[2,3,*] Haili Song,[1] Jing Wang,[1] Guangqin Li,[1] Corey E. Frank,[4] Lu Ma,[5] Mark Croft,[6] Yonggang Wang,[7] Martha Greenblatt,[4] Dao-Xin Yao,[2,†] and Man-Rong Li[1,‡]

[1] *Key Laboratory of Bioinorganic and Synthetic Chemistry of Ministry of Education, School of Chemistry, Sun Yat-Sen University, Guangzhou 510275, China.*

[2] *State Key Laboratory of Optoelectronic Materials and Technologies, School of Physics, Sun Yat-Sen University, Guangzhou 510275, China.*

[3] *College of science, Hangzhou Dianzi University, Hangzhou 310018, China.*

[4] *Department of Chemistry and Chemical Biology, Rutgers, The State University of New Jersey, 123 Bevier Road, Piscataway, New Jersey 08854, USA*

[5] *Brookhaven National Laboratory, National Synchrotron Light Source II, Bldg. 74, P.O. Box 5000, Upton, NY 11973-5000*

[6] *Department of Physics and Astronomy, Rutgers, The State University of New Jersey, 136 Frelinghuysen Road, Piscataway, New Jersey, 08854, USA*

[7] *Center for High Pressure Science and Technology Advanced Research (HPSTAR), Beijing 100094, China*

*Corresponding author: zengyj@hdu.edu.cn
†yaodaox@mail.sysu.edu.cn
‡limanrong@mail.sysu.edu.cn

# Contents

# 1. Supplementary Notes

## 1.1 UV-Vis DRS analysis

Figure S5 displays the UV-Vis DRS spectra of CTO, CTON-700, CTON-750 and CTON-800. The absorption edge of samples gradually shifts to longer wavelengths as the preparation temperature increases up to 800 °C. The color evolution shown in Fig. 1(a) echoes to the temperature-dependent shift trend of the band gap determined by the Kubelka-Munk function [1,2]. The CTO synthesized in this work shows direct band gap energy ($E_g$) of 3.22 eV, which is consistence with its insulator nature [1]. With the increment of N content in $CoTa_2O_{6-x}N_{2x/3}\square_{x/3}$, the band gap narrows to 3.02 eV as seen in Fig. S5(b). The $E_g$ variation in $CoTa_2O_{6-x}N_{2x/3}\square_{x/3}$ is comparable to previous discoveries in $RbLaTa_2O_{6.77}N_{0.15}$ ($E_g$ = 2.09 eV) and $RbLaTa_2O_7$ ($E_g$ = 4.26 eV), where DFT calculations also illustrated that the narrowing of band gaps is mainly due to the upward shift of the valence band contributed by N $2p$ orbital [3].

## 1.2 XANES and XPS analysis

*Co K-edges:* The near edge features at the K edges of $3d$ row transition metal, T($3d$), compounds are due to transitions from the $1s$ to $4p$ states of the transition metal, combined with a step feature for the continuum onset [4,5]. The systematic T($3d$)-K main-edge energy shift to higher energy with increasing valence can serve as indicator of charge transfer in such compounds. The Co-K main-edges of CTO and CTON-800 are compared to various standard oxide compound spectra (with octahedral T-O coordination) in Fig. S8(a). Here the CoO standard involves multiple edge sharing octahedra [6,7] and leads to additional splittings in the main-edge $p$-features, whereas the corner sharing octahedra in the perovskite based $La_2CoVO_6$, $LaCoO_3$ and $SrCoO_{3-\delta}$ standards lead to a sharper $4p$ peak feature. The clear chemical shift, to higher energy, of the steeply rising edge onset between the ~ $Co^{2+}$, ~$Co^{3+}$, and ~$Co^{4+}$ standards is indicated in the Fig. S8(a). The low-energy of the edge onset for the CTO and CTON-800 spectrums evidence an $Co^{2+}$ assignment in the materials. Parenthetically, it is worth noting that the edge sharing in the CTO based compounds leads to split 4p features as in the CoO standard.

*Ta-L$_{3,2}$ edges*: The L$_{2,3}$ edges of 5d-transition metals are dominated by very intense "white line" (WL) features (Fig. S8(b)) due to dipole transitions into final 5d hole states [8-10] which vary systematically and dramatically with the 5d-orbital electron (hole) count. This is illustrated by comparing the octahedrally coordinated standard $5d^0$, $Ca_2MnTaO_6$ (top) and $5d^1$, $Ba_2MnReO_6$ (bottom) spectra in Fig. S8(b), where the distinct bimodal A/B WL-peak-structures manifest the conspicuous increase of relative A-feature intensity with increasing d-hole count. Here, the L$_2$-edges and L$_3$-edges have been superimposed in energy to normally align the B-features.

An oft used schematic diagram [11-13] of d-orbital splittings due to octahedral ligand crystalline electric field (CEF) and the relativistic atomic spin orbit (SO) effects is shown in Fig. S8(c). Figure S8(c) (left) motivates the association of the A-feature with holes in the (CEF) lower energy $t_{2g}$ orbitals (6X degenerate) and the B-feature with holes in the higher energy $e_g$ orbitals (4X degenerate) [8-10]. Thus, the dramatic A feature intensity variation, between the $5d^0$ and $5d^1$, is directly associated with the 6 to 5 d-hole final state change. It should be noted that in the $5d^0$ to $5d^1$ standard spectra, the A-feature intensity decrease is noticeably larger. Referring to the schematic orbital splitting diagram in Fig. S8(c), the SO orbital splitting of the d-states into $J = 3/2$ and $5/2$ orbitals should be noted. Moreover, in the center panel, where both SO and CEF orbital splittings are combined, one should note that the lowest lying CEF-$t_{2g}$ orbitals have $J = 3/2$ character. Thus, the L$_2$ edge, with its dipole selection rule dictated $d^{3/2}$ final states is consistent with the observed enhanced sensitivity of the A-feature to changes in the d-occupancy in this range.

Turning to the CTO and CTON-800 spectra in the center of Fig. S8(b), on can see that the relative A- to B-feature intensity is clearly consistent with the ~$Ta^{5+}$ configuration in both CTO and CTON-800.

The surface compositions, chemical states, and chemical bonds of CTO, CTON-700, CTON-750 and CTON-800 were determined by XPS. As shown in Fig. S9(a), the Co $2p_{3/2}$ and $2p_{1/2}$ peaks are observed at ~780.88 and ~796.78 eV, respectively [14,15]. It shows a very slight shift as nitrogen content increases, indicating the stable existence of $Co^{2+}$ oxidation state and ruling out the existence of any possibly reduced Co impurity. Figure S9(b) shows the O $1s$ spectra, the major peaks at 530.18~530.58 eV are typical metal-oxygen bonds. The peaks at 531.58 eV correspond to oxygen vacancies due to the nitridation of CTO, and the peaks at 532.03 eV are attributed to surface adsorbed water molecules [16,17]. In Fig. S9(c), the peaks around 396.18 eV are assigned to the N $1s$, indicating the formation of Co/Ta-N bonds enhanced by higher N substitution. The Ta $4p_{3/2}$ peaks located at 404.48 eV indicates $Ta^{5+}$ [18-21]. The peak at 399.38 eV suggests the presence of N-H bonds, which can be attributed to either the incomplete reaction of $NH_3$ or adsorbed atmospheric water [22].

**1.3 FT-IR analysis**

The interactions among chemical bonds in $CoTa_2O_{6-x}N_{2x/3}\square_{x/3}$ have been further investigated by FT-IR spectra as shown in Fig. S10. As can be seen, all samples exhibit similar spectra, the signals at 3450 and 1637 cm$^{-1}$ are associated with the O-H stretching vibrations and H-O-H bending vibrations of surface adsorption $H_2O$ molecules, respectively [23]. The bands at 870 and 640 cm$^{-1}$ are assigned to Ta-(O/N) stretching and bending vibration of Ta-(O/N)-Ta, respectively [24,25]. The vibration bands at 476 cm$^{-1}$ correspond to stretching modes of Co-(O/N) [26,27]. By comparison, the Co-(O/N) vibration strengthens with increasing N content, but that of Ta-(O/N) is unobvious. This is probably because the atomic mass of tantalum is larger than that of cobalt, the slight N amount is insufficient to cause stretching vibration of Ta-(O/N). In short, the FT-IR results are well consistent with the XPS and XANES results.

## 2. Supplementary Figures

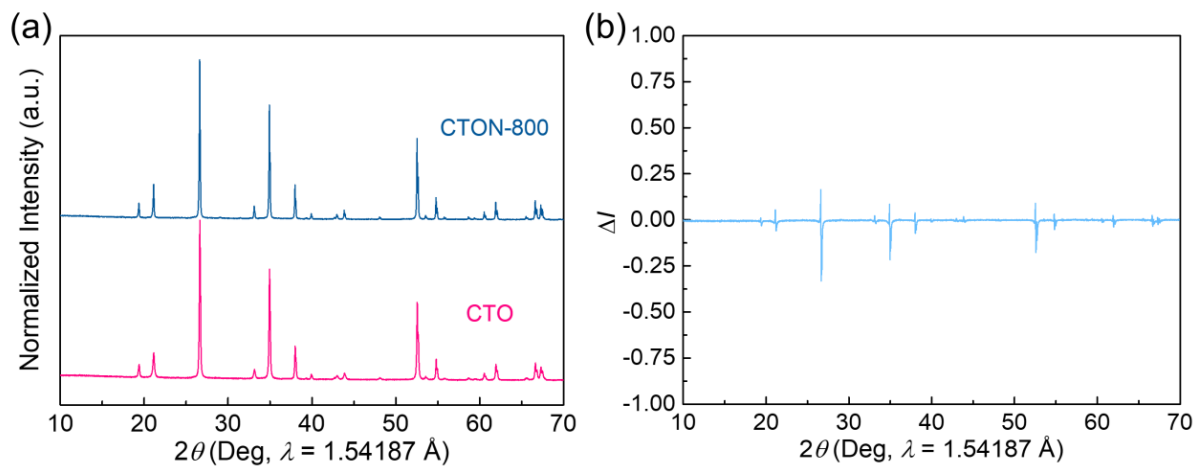

**Fig. S1** (a) Comparison of peak intensity between normalized XRD patterns of CTO and CTON-800. (b) The intensity difference ($\Delta I = I$(CTON-850)-$I$(CTO)) between normalized XRD patterns of CTO and CTON-800.

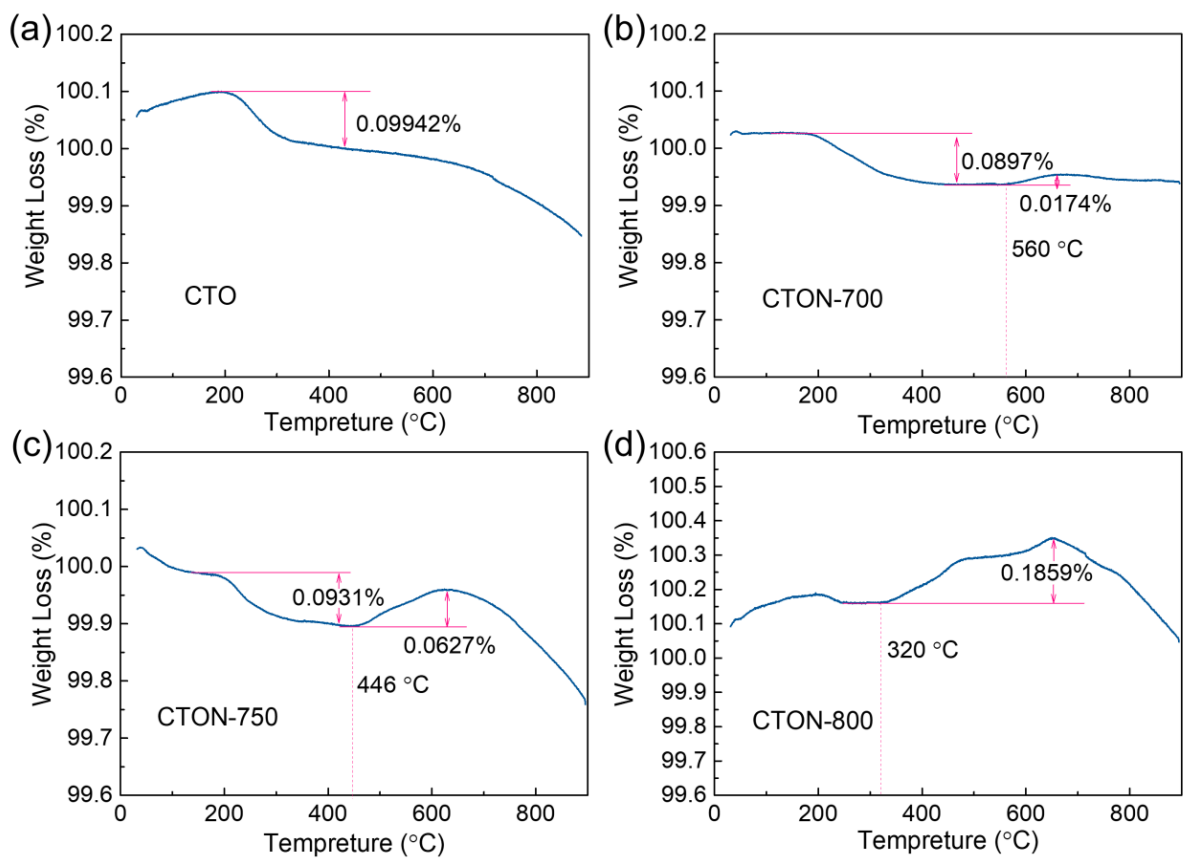

**Fig. S2** TGA curves of (a) CTO, (b) CTON-700, (c) CTON-750 and (d) CTON-800 in air (flowing rate of 50 mL/min) from 30 to 900 °C at a heating rate of 10 °C min$^{-1}$.

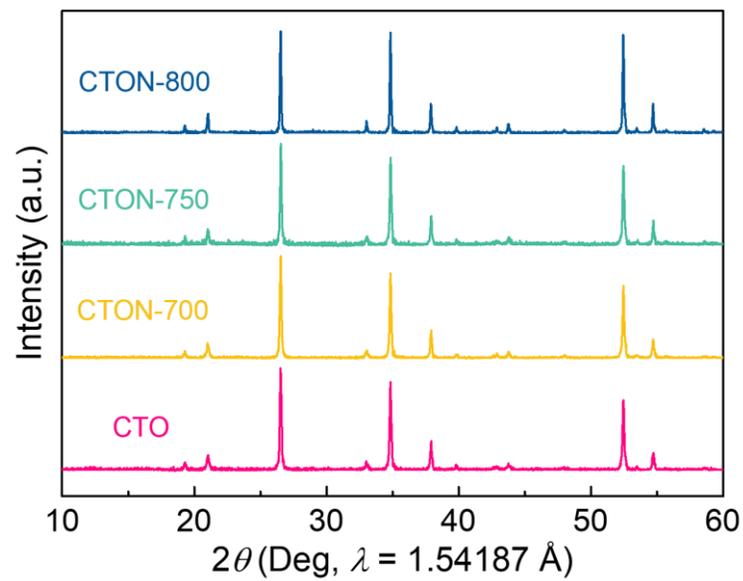

**Fig. S3** XRD patterns for $CoTa_2O_{6-x}N_x$ samples after TGA measurements.

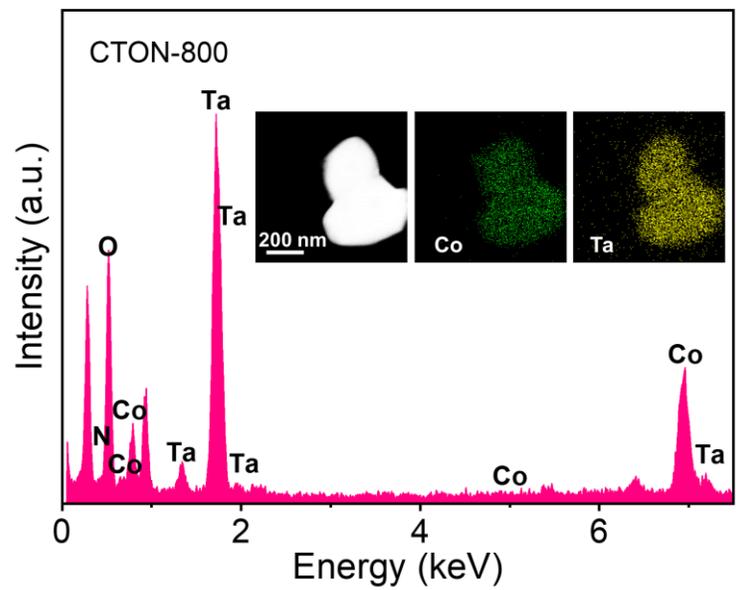

**Fig. S4** TEM-EDS spectra of CTON-800. Insert shows the TEM-EDS mapping of a representative particle of CTON-800.

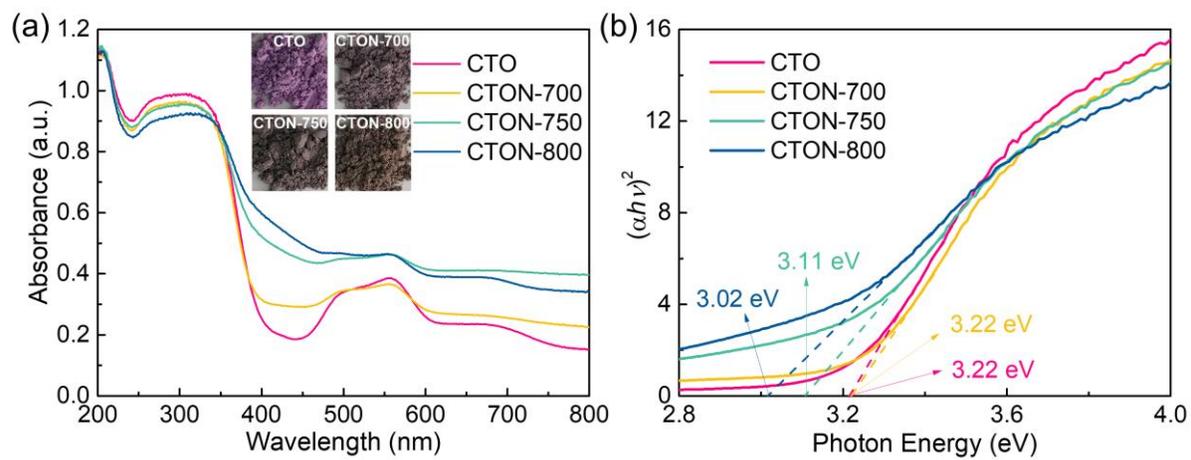

**Fig. S5** (a) UV-Vis DRS and (b) Plot of (α*hv*)² *vs. hv* for band gap of CTO, CTON-700, CTON-750 and CTON-800.

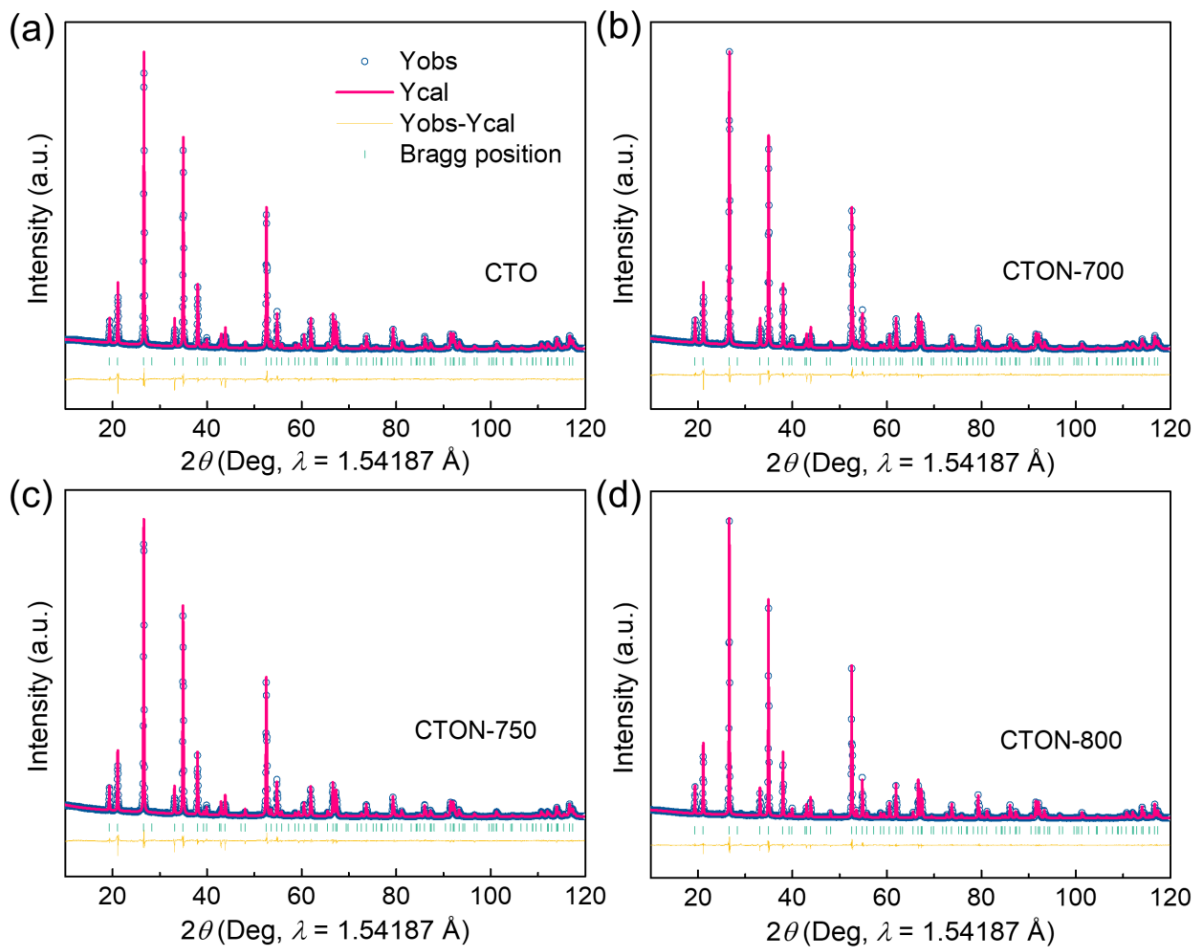

**Fig. S6** Rietveld refinement of the XRD patterns of (a) CTO; (b) CTON-700; (c) CTON-750; (d) CTON-800.

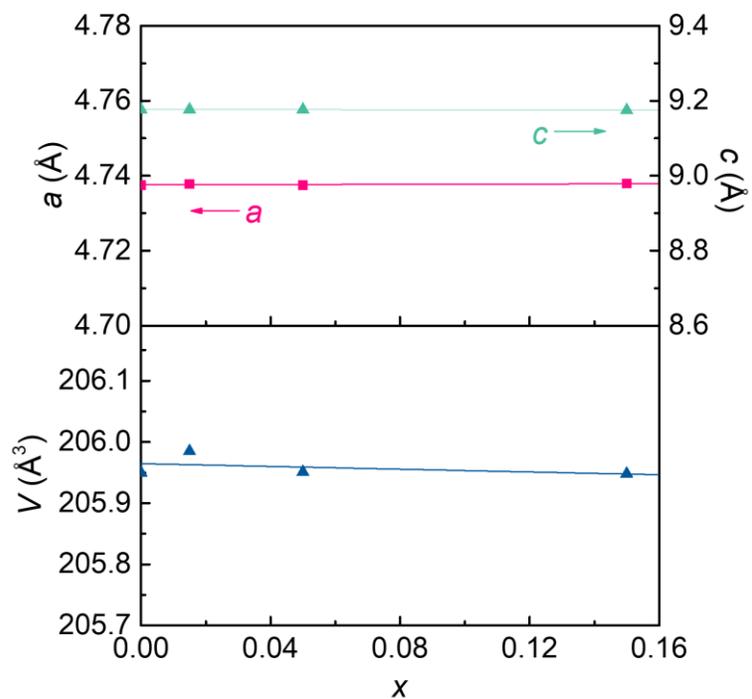

**Fig. S7** $x$-dependent variation of the tetragonal lattice parameter $a$ (Å), $c$ (Å), and $V$(Å$^3$) in CoTa$_2$O$_{6-x}$N$_{2x/3}$□$_{x/3}$ refined from XRD data.

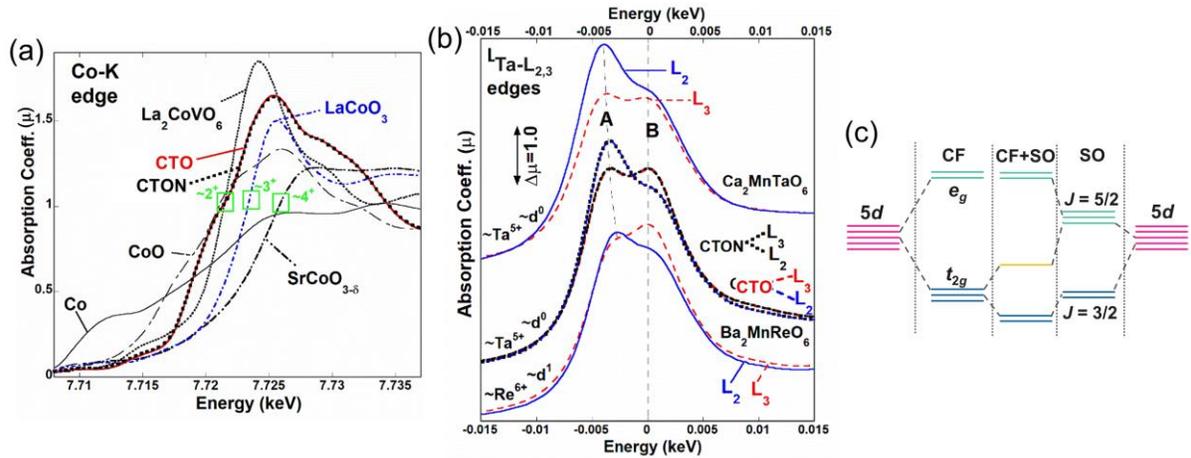

**Fig. S8** (a) The Co-K main edge of CTO and CTON-800 compared with those of a series of standard spectra: ~$Co^{4+}$, $SrCoO_{3-\delta}$; ~$Co^{3+}$ $LaCoO_3$; ~$Co^{2+}$, CoO; and elemental Co. (b) The superimposed $L_{2,3}$-edge spectra for: the octahedrally coordinated standard $5d^1$, $Ca_2MnTaO_6$ (top) and $5d^1$, $Ba_2MnReO_6$ (bottom) compounds; for comparison to the CTO and CTON-800 (middle) compounds. Here, the $L_2$-edges and $L_3$-edges have been superimposed in energy to normally align with the $e_g$-hole-related the B-features. It is worth noting that the $Ca_2MnTaO_6$ compound is perovskite-*B*-site disordered and is broadened. (c) A Schematic of the *d*-shell orbital diagram with: spherical symmetry shown on the extreme right and left; the octahedral crystal field (CF) splitting into $e_g$ and $t_{2g}$ orbitals, labeled CF; the spin orbit (SO) splitting of $J = 3/2$ and $J = 5/2$ orbitals, labeled SO; and the combined CF+SO regime center. Note, the systematic/dramatic decrease of the intensity of the $t_{2g}$-hole-related A-feature, relative to the B-feature, as the $t_{2g}$ *d*-electron-count varies between $5d^0$ and $5d^1$ ($t_{2g}$ *d*-hole-count between 6 and 5) in the standard spectra. Importantly, this is A-feature decrease is especially dramatic in the $L_2$-edge spectra with the strong spin orbit (SO) effect in these $5d$ materials. As noted in the main portion of the text the relative A- to B-feature intensities in the CTO and CTON-800 spectra clearly support the ~$5d^1$, $Ta^{5+}$ assignment for both compounds.

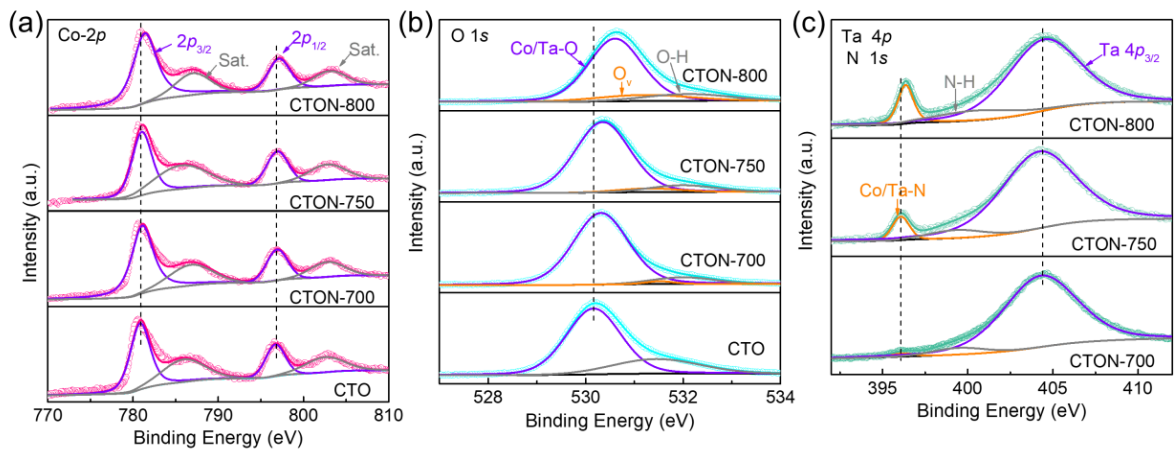

**Fig. S9** High-resolution XPS spectra of (a) Co 2*p*, (b) O 1*s*, (c) Ta 4*p* and N 1*s* of CTO, CTON-700, CTON-750 and CTON-800.

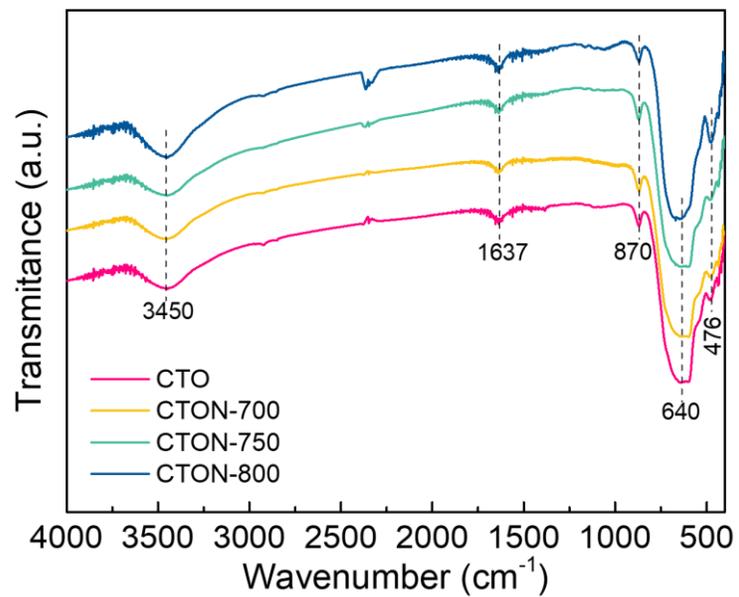

**Fig. S10** FT-IR spectra of CTO, CTON-700, CTON-750 and CTON-800.

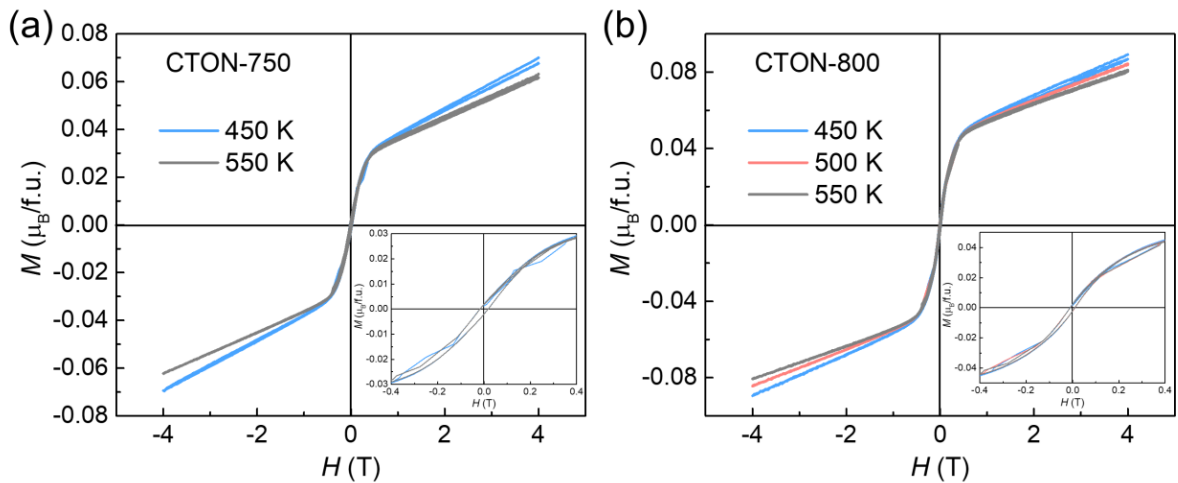

**Fig. S11** Isothermal magnetization curves of $CoTa_2O_{6-x}N_{2x/3}\square_{x/3}$ measured at 450, 500, and 550 K between -4 and 4 T. (a) CTON-750; (b) CTON-800. The insets show the enlarged area between -0.4 and 0.4 T.

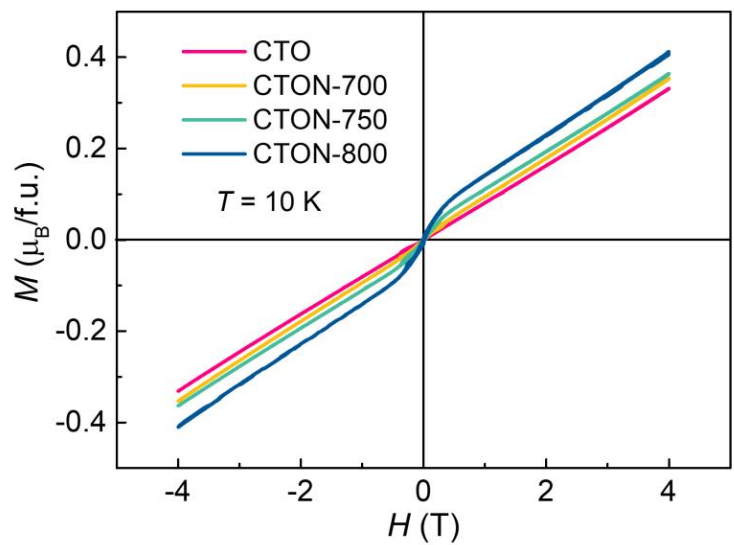

**Fig. S12** Isothermal magnetization curves for CoTa$_2$O$_{6-x}$N$_{2x/3}$□$_{x/3}$ samples measured at 10 K under a magnetic field ranging from -4 to 4 T.

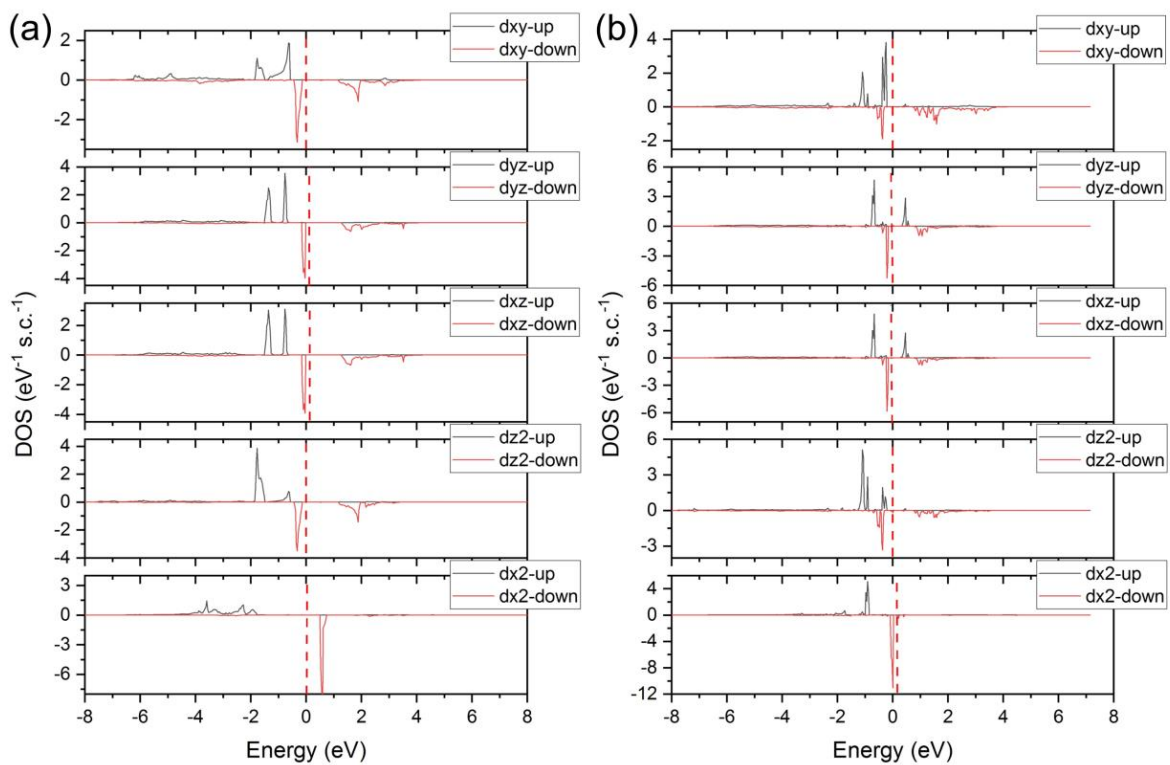

**Fig. S13** (a) The PDOS of *d* orbitals of Co atom in the corner of bulk CTO. The unit of the vertical coordinate is number of states per eV per supercell. (b) The PDOS of *d* orbitals of Co atom with $CoO_5N$ coordination and Co-N bond not being on *a-b* plane ($Co_{3,a}$ in Figure 6 (a)). The magnetic moment of Co atom in (a) is 2.55 $\mu_B$, and that in (b) is 1.17 $\mu_B$. The reduced magnetic moment in (b) comes from the fact that the spin-up $d_{yz}$ and $d_{xz}$ orbitals move higher into the conduction band and are not fully occupied, and the spin-down $d_{x2-y2}$ orbital moves into the valence band and becomes occupied. This change is a consequence of the change of crystal field of Co atom.

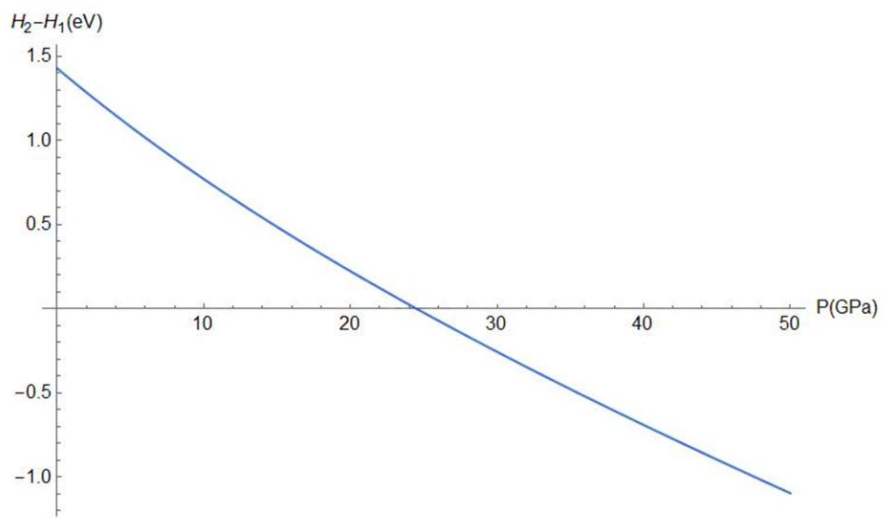

**Fig. S14** The pressure dependence of the relative enthalpies between phase 2 ($V < 190$ Å$^3$) and phase 1 ($V > 190$ Å$^3$) of CTO.

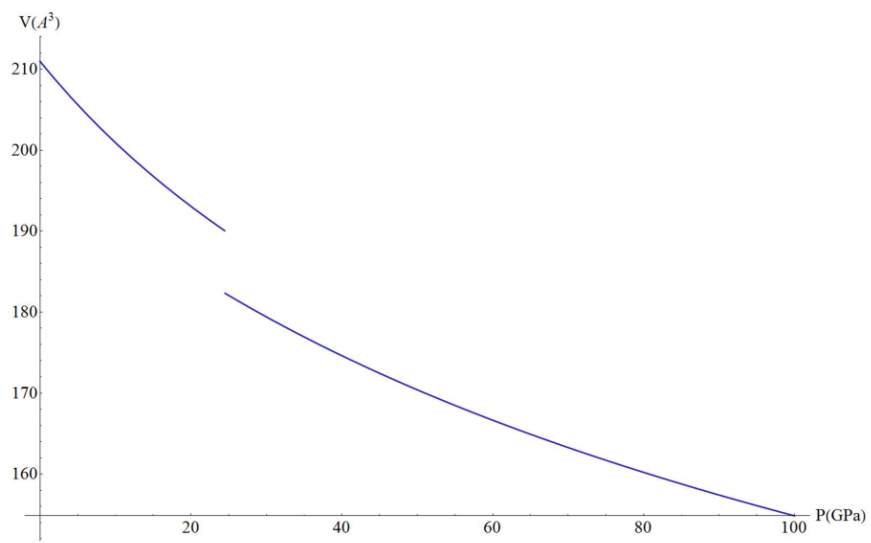

**Fig. S15** The pressure dependence of unit cell volume of CTO. The volume has a discontinuity at the transition pressure.

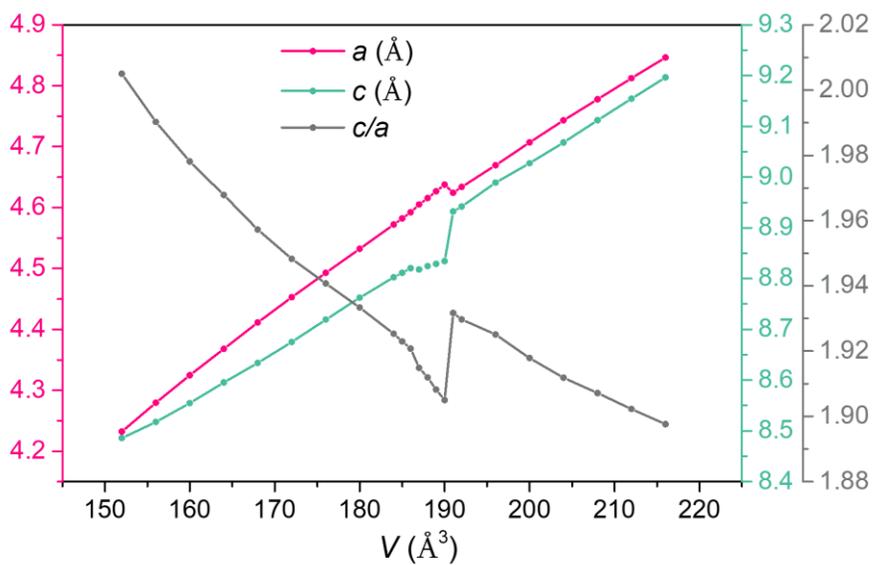

**Fig. S16** Evolution of lattice constants *a*, *c* and *c/a* as a function of the unit cell volume of CTO. As the pressure increases (with volume decreasing), to the phase transition point, there is a sudden decrease in *c* and increase in *a*, leading to discontinuous *c/a* at the phase boundary.

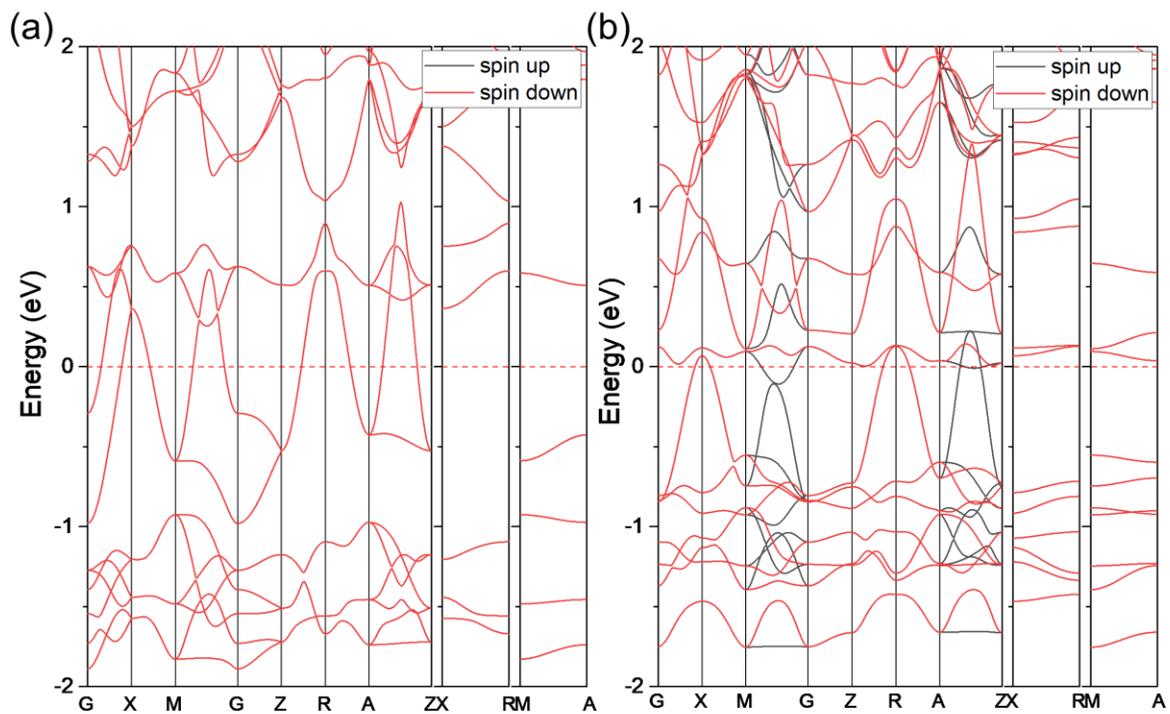

**Fig. S17** The calculated spin polarized band structure of CTO with unit cell (a) volume 168 Å$^3$, which is in phase 2, and (b) volume 188 Å$^3$, which lies on the phase boundary, respectively. Both are metallic, however, in (b) the *k*-dependent spin splitting is preserved, while in (a) this feature disappears due to the disappearance of the magnetic moment of cobalt.

## 3. Supplementary Tables

**Table S1** Structure refinements results of fractional atomic coordinates, occupancies and isotropic atomic displacement parameters of CTO, CTON-700, CTON-750 and CTON-800.

|  | CTO | CTON-700 | CTON-750 | CTON-800 |
|---|---|---|---|---|
| $a/\text{Å}$ | 4.7375(4) | 4.7378(4) | 4.73748(4) | 4.73789(2) |
| $c/\text{Å}$ | 9.1764(9) | 9.1765(9) | 9.17626(9) | 9.17457(4) |
| $V/\text{Å}^3$ | 205.95(5) | 205.988(5) | 205.947(5) | 205.948(2) |
| Co1 2$a$ (0, 0, 0) | | | | |
| $B_{iso}/\text{Å}^2$ | 0.208(10) | 0.262(10) | 0.276(10) | 0.278(7) |
| Ta1 4$e$ (0, 0, $z$) | | | | |
| $z$ | 0.33181(7) | 0.33197(6) | 0.33195(7) | 0.33161(5) |
| $B_{iso}/\text{Å}^2$ | 0.208(10) | 0.262(10) | 0.276(10) | 0.278(7) |
| O1/N1 4$f$ ($x$, $y$, 0) [a] | | | | |
| $x$ | 0.2803(6) | 0.2798(6) | 0.2787(6) | 0.2997(5) |
| $y$ | 0.2803(6) | 0.2798(6) | 0.2787(6) | 0.2997(5) |
| Occ. | 1 | 0.9925/0.0050 | 0.9750/0.0150 | 0.9250/0.0500 |
| $B_{iso}/\text{Å}^2$ | 0.208(10) | 0.262(10) | 0.276(10) | 0.278(7) |
| O2 8$j$ ($x$, $y$, $z$) | | | | |
| $x$ | 0.3155(3) | 0.3155(3) | 0.3163(3) | 0.3050(2) |
| $y$ | 0.3155(3) | 0.3155(3) | 0.3163(3) | 0.3050(2) |
| $z$ | 0.3350(5) | 0.3348(5) | 0.3366(5) | 0.3296(4) |
| Occ. | 1 | 1 | 1 | 1 |
| $B_{iso}/\text{Å}^2$ | 0.208(10) | 0.262(10) | 0.276(10) | 0.278(7) |
| $R_{wp}$ (%) | 5.87 | 5.72 | 5.90 | 4.35 |
| $R_p$ (%) | 7.16 | 7.05 | 7.11 | 5.93 |

[a] The O/N was refined by disorder distribution at 4$f$ ($x$, $y$, 0) site and the occupancy was fixed according to $x$ in the formula.

**Table S2** Selected bond lengths (Å) and bond angles (deg) for CTO, CTON-700, CTON-750 and CTON-800.

|  | CTO | CTON-700 | CTON-750 | CTON-800 |
|---|---|---|---|---|
| Co1-(O1/N1) (×2) | 1.878(4) | 1.875(3) | 1.867(4) | 2.009(3) |
| Co1-O2 (×4) | 1.953(4) | 1.955(4) | 1.940(4) | 2.027(3) |
| <Co1-O/N> | 1.928(4) | 1.928(4) | 1.916(5) | 2.027(3) |
| Ta1-(O1/N1) (×2) | 2.133(3) | 2.134(3) | 2.139(3) | 2.046(16) |
| Ta1-O2 (×2) | 2.115(15) | 2.114(15) | 2.120(15) | 2.044(15) |
| Ta1-O2 (×2) | 1.968(4) | 1.968(4) | 1.977(4) | 1.974(3) |
| <Ta1-O/N> | 2.072(7) | 2.072(7) | 2.079(7) | 2.021(11) |
| (O1/N1)-Co1-(O1/N1) | 180.0(0) | 180.0(0) | 180.0(0) | 180.0(0) |
| (O1/N1)-Co1-O2 | 90.0(8) | 90.0(7) | 90.0(8) | 90.0(6) |
| O2-Co1-O2 | 78.5(19) | 78.4(19) | 78.8(19) | 79.8(15) |
|  | 101.5(19) | 101.6(19) | 101.2(19) | 100.2(15) |
|  | 180.0(0) | 180.0(0) | 180.0(0) | 180.0(0) |
| (O1/N1)-Ta1-(O1/N1) | 87.3(14) | 87.5(12) | 87.8(14) | 81.9(11) |
| (O1/N1)-Ta1-O2 | 97.5(12) | 97.4(11) | 97.6(11) | 97.6(10) |
|  | 89.4(14) | 89.5(13) | 89.2(13) | 90.4(11) |
|  | 175.3(12) | 175.2(11) | 174.6(12) | 179.5(10) |
| O2-Ta1-O2 | 90.6(12) | 90.6(12) | 90.9(12) | 89.6(11) |
|  | 77.8(19) | 77.8(19) | 77.0(18) | 82.9(16) |
|  | 178.4(3) | 178.5(3) | 177.7(3) | 179.0(3) |